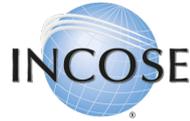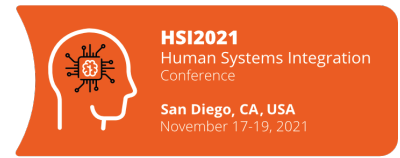

# Socioergonomics:

## A few clarifications on the Technology-Organizations-People Tryptic

Guy André Boy, Ph.D., INCOSE Fellow
Paris Saclay University (CentraleSupélec) & ESTIA Institute of Technology
9 rue Joliot Curie, 91190 Gif-sur-Yvette, France
+33 6 73 11 79 38
guy-andre.boy@centralesupelec.fr

**Abstract**. This position paper introduces and coins the term socioergonomics, considered as a sociological, ontological, and methodological support to human systems integration (HSI). It describes the evolution of ergonomics from early physiological to psychological to contemporary social sciences approaches supporting Industry 4.0 sociotechnical systems engineering. It presents a Technology Readiness Levels (TRLs) extension to Organizational Readiness Levels (ORLs) and a departure toward a socioergonomics approach that includes systemic properties such as flexibility, separability, and emergent social facts.

## Introduction

For the last two years, International Council on Systems Engineering (INCOSE) Human Systems Integration (HSI) working group is working on a 3-pager chapter that synthesizes HSI. We came up with the following diagram (Figure 1), where TOP Model (Technology-Organizations-People; Boy, 2020) is the central concurrent design and management process support within an environment. Indeed, HSI cannot be thought without concurrent and incremental design of technology, organizations, and people's activities, and consequently jobs. The environment is where the TOP Model is defined (e.g., health care environment, airspace, battlefield). This should be seen from various perspectives: safety; competence and professionalism; sustainability; habitability; occupational health; social, cultural, and organizational factors; training; HSI planning; human factors engineering; workforce planning; integrated logistics support and maintenance. These perspectives require four core disciplines and practices: Systems Engineering; Human Factors and Ergonomics (HFE); Information Technology; and the Operational Domain at stake.

The HFE part can be investigated from a large variety of interests that include physiological, psychological, and sociological factors. This paper focuses on HSI sociological factors, and proposes a viewpoint, coined as "socioergonomics". The main purpose of this definitional contribution is to provide an articulated set of concepts toward an ontology useful for HSI development from a societal, cultural, and organizational perspective (Norman, 2019). Instead of considering single-agent cognitive modeling from an integrative approach going from the human nervous system to measurable indicators like what neuroergonomics proposes (Parasuraman, 2003), socioergonomics considers a multi-agent socio-cognitive modeling approach that is based on organizational knowledge of the world influencing agents involved in, and reciprocally. This approach requires an integrative definition of what a system is about and the various relationships this system has with the various systems surrounding it. In other words, we will better define what is inside an agent or system by mastering what influences it from the outside. Note that the meaning of the term "agent" used in artificial intelligence is very similar to what "system" means in systems engineering.

Socioergonomics is strongly related to macroergonomics that provides knowledge and methods toward the effectiveness and performance of work systems (Hendrick, 2002). Macroergonomics, a systemic branch of HFE, considers the organizational and sociotechnical context of work activities and processes. It emphasizes sociotechnical systems (STSs). STS integration is an instance of HSI. For example, the SEIPS (Systems Engineering Initiative for Patient Safety) model of work system and patient safety is a sociotechnical approach successfully applied in healthcare research and practice. Carayon and her colleagues claimed that healthcare systems and processes need to be systematically redesigned. They provided several macroergonomic approaches, principles, and methods for healthcare system redesign (Carayon et al., 2013).

Why should not we keep the term macroergonomics instead of inventing a new term such as socioergonomics? Both refer to sociotechnical systems. First, socioergonomics is more anchored into sociology than psychology. Psychology is devoted to study the mind of an individual (i.e., the single agent approach). Instead, sociology extends the study of individuals toward society (i.e., the multi-agent approach). Sociology is "a social science that studies human societies, their interactions, and the processes that preserve and change them. It does this by examining the dynamics of constituent parts of societies such as institutions, communities, populations, and gender, racial, or age groups." (https://www.britannica.com/topic/sociology: Encyclopædia Britannica, 2021). Sociology is devoted to the study of communities and organizations to better understand their internal and external interdependencies and properties. In systems engineering, we would talk about Systems of Systems (SoS). As a matter of fact, SoS representations should nicely support sociological studies. This is the reason why socioergonomics is currently developed to support HSI by providing not only a vocabulary for talking about it, but also meaningful societal representations and patterns for HSI analysis, design, and evaluation.

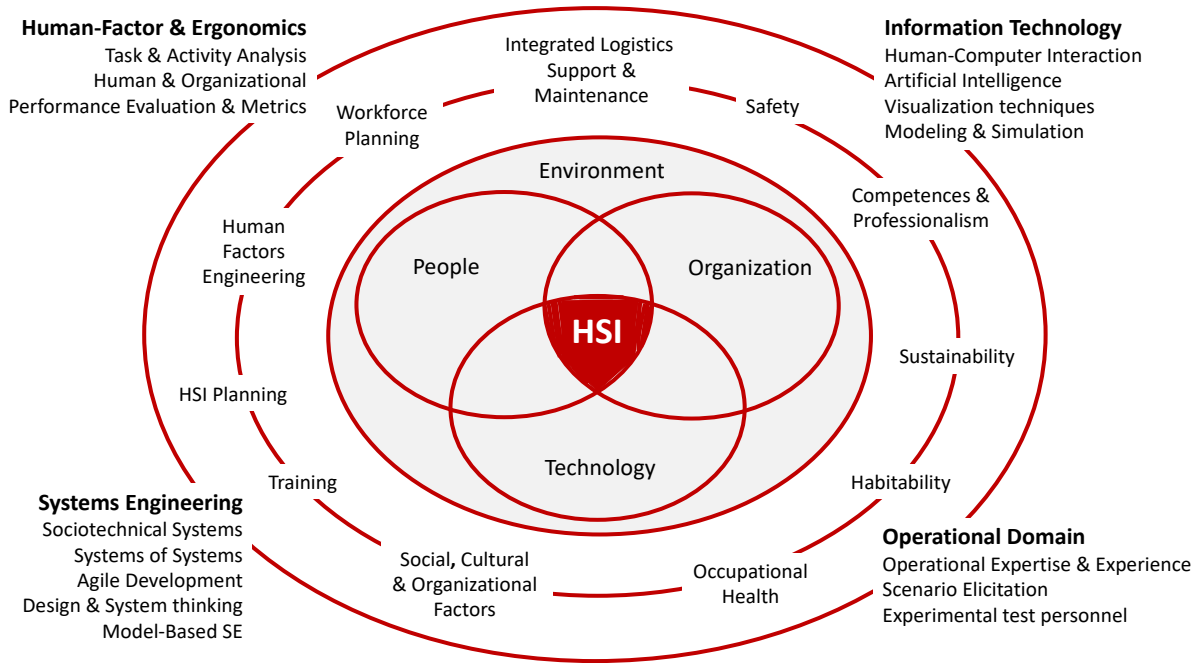

Figure 1. HSI as a product and a process within three layers going from core disciplines and practices to industrial development perspectives, to the TOP Model at the center. This diagram comes from the Human Systems Integration Chapter for INCOSE SE Handbook, 5th Edition, Work in progress of INCOSE HSI Working Group (to appear in 2022).

## Evolution of Ergonomics

The International Ergonomics Association (IEA) defines *ergonomics as* "the science of work", derived from the Greek *ergon* (work) and *nomos* (laws). "Ergonomics (or human factors) is the scientific discipline concerned with the understanding of interactions among humans and other elements of a system, and the profession that applies theory, principles, data, and methods to design in order to optimize human well-being and overall system performance".

In practice, ergonomics has been a corrective more than a design discipline. Human-Centered Design (HCD) started to become credible at the turn of the 21st century when Human-In-The-Loop Simulations (HITLS) became increasingly tangible both physically and figuratively (Boy, 2016, 2020). HITLS were used before but in very specific ways, for isolated subsystems of a system. Today, we can develop virtual prototypes of the whole system, considered as a system of systems (Popper, Bankes, Callaway, & DeLaurentis, 2004). In the meantime, Human-Computer Interaction (HCI) and Artificial Intelligence (AI) contributed to enhance HITLS toward more holistic HCD. HITLS enables us to further develop the relationship between HCD teams, as anthropologists, and their object of study as a particular instance of the relationship between knowing and doing, interpreting, and using, symbolic mastery and practical mastery of a system. Bourdieu developed the "logic of practice" to this end in his anthropological investigations (Bourdieu, 1992).

As already said, the paradigm shift from single agent to multi-agent systems contributed to the cognitive engineering shift from cognition to socio-cognition. For a long time, ergonomics was dominated by physiologists (after World War 2) and psychologists (since the early 1980s during personal computers development and massive software expansion in our everyday lives). Today, information and transportation technologies enable us to be more interconnected. They also increased and modified the nature of societal complexity. This is the reason why sociology, anthropology and social sciences in general are emerging disciplines in ergonomics. We cannot explore human activities in the real-world without considering cultural issues and practices. It is interesting to notice, for example, how various countries developed very different solutions to current COVID-19 pandemic problems. Indeed, facing unexpected situations requires using deeper knowledge, skills and ways of doing things. In these situations, people use socially admitted and organizationally driven approaches, sometimes based on educated common sense (Boy, 2021). Ergonomics almost always ends up with procedure-based solutions; socioergonomics requires more problem-solving approaches that require collaboration using several types of knowledge and skills.

Summing up, ergonomics has evolved, going from *inside-out* approaches based on technology-centered engineering (i.e., building core insider technology first) followed by user interface design, adapting people to machines, to *outside-in* approaches based on HCD (i.e., establishing the TOP-centered purpose first) followed by agile technology development, co-adapting technology, organizations, and people. HCD requires participation, collaboration, and coordination of expert stakeholders, as well as development and refinement of socioergonomics methods and tools.

## Socioergonomics in Industry 4.0

Since the turn of 21st century, we have entered an increasingly digital world, where virtual HCD (VHCD) is concretely used in engineering design and systems engineering. VHCD enables to observe people's activity in virtually simulated possible future systems. Agile development is then possible. However, virtual environments bring to the front tangibility issues. Tangibility of a system addresses the grasp of its physical aspects and cognitive aspects (i.e., its meaning). Consequently, methods and tools need to be developed to support progressive tangibilization of virtual prototypes toward concrete products (Figure 2). This tangibilization process involves various kinds of stakeholders, and therefore requires us to address concepts and metrics such as trust, collaboration, and coordination. Digitalization does not concern only interaction between people and increasingly digital machines, but also between people through this type of smart machines. Consequently, both technological maturity and organizational maturity need to be assessed.

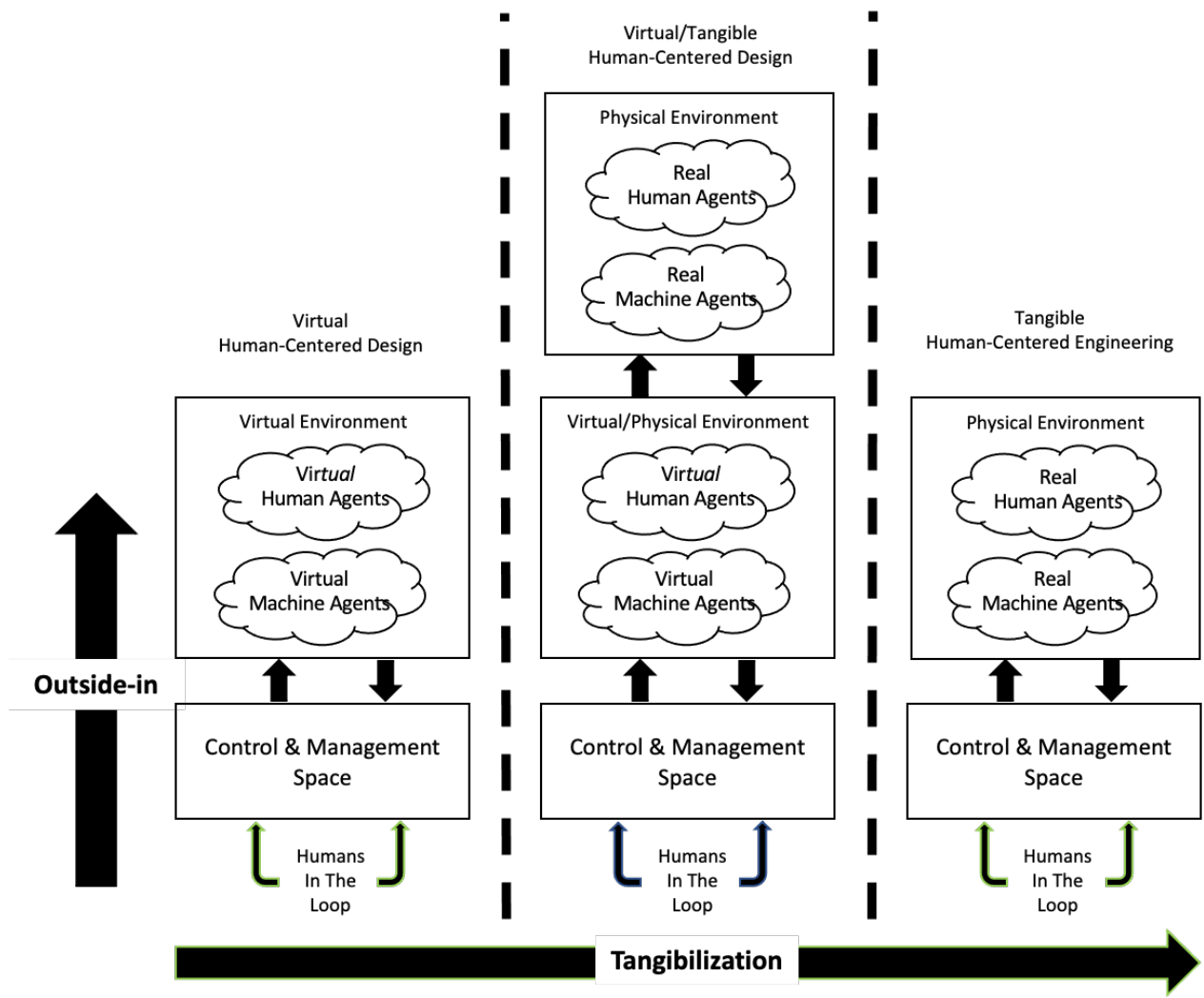

Figure 2. Tangibilization process from VHCD to a tangible sociotechnical system product.

NASA Technology Readiness Levels (TRLs) developed on a scale from 1 to 9 led to metrics adapted to industrial needs for the assessment of technology maturity along the life cycle of a system (https://www.nasa.gov/directorates/heo/scan/engineering/technology/technology_readiness_level).

Socioergonomics needs to address organizational readiness in the same way. Organizational performance depends on the way people's jobs and machines are set up. Today, machines include a fair amount of AI, providing them with increasingly complex interactive behaviors. Organizational Readiness Levels (ORLs) can be developed along with the TRLs. ORLs are correlated with tangibility evolution to be observed during the development of a sociotechnical system from the first ideas to its delivery (Figure 2).

From the beginning of the life cycle of a system, we take an outside-in approach going from purpose to means, considering the control and management space as an element of the overall virtual system, which is incrementally tangibilized. Once the system is fully tangibilized, personnel training and potential refinement of the control and management space can be further developed until a satisfactory HSI is stabilized (taking the old inside-out approach). A first cut of ORLs levels is provided on Table 1.

Table 1. Organizational Readiness Levels.

| ORL-0 | About first principles where potential organizational models are explored |
|---|---|
| ORL-1 | Goal-oriented research that requires making choices from first principles to practical fully digital organizational setups |
| ORL-2 | Proof of principle development and active R&D is started in a virtual environment |
| ORL-3 | Virtual agile organizational prototype development and first HITLS (virtual HCD) |
| ORL-4 | Proof of organizational concept development using concrete scenario-based design from fully virtual to more tangible environments |
| ORL-5 | Assessing organization capability in terms of authority sharing (responsibility, accountability, and control), trust, collaboration and coordination, for example |
| ORL-6 | Real-world use-case tests in a wider variety of situations – tangibilization continues |
| ORL-7 | Practical integration with respect to criteria such as safety, efficiency and comfort, at various levels of granularity of the organization – tangibilization continues |
| ORL-8 | Readiness for effective implementation on a real site (fully tangible) based on personnel feedback for deployment approval |
| ORL-9 | Deployment involving both personnel and real machines |

## *Toward on Ontology of Socioergonomics*

We have become accustomed to confusing the notion of system with that of machine. In fact, a system can represent an entity that is natural and/or artificial. Medical doctors talk about the cardiovascular system or the neural system of a human being. Lawyers talk about the legal system of a country. Politicians talk about a centralized or federal state system. It is sometimes difficult to extend the concept of system to represent humans. Nevertheless, a human being can be represented as a system composed of other systems, which may be natural (e.g., the human body includes organs, limbs and speech) or artificial (e.g., a pacemaker, a prosthesis). Symmetrically, a machine system may include human systems (e.g., an aircraft includes pilots and passengers). Therefore, the notion of system can be considered as a representation (Figure 3).

From a teleological point of view, a system can be described by a structure and a function, which both have a role, a context of validity and resources. A resource can be physical and/or cognitive and can typically be a human and/or machine system. Consequently, a system is recursively defined as a system of systems (i.e., a system includes other sub-systems and belongs to bigger systems). Consequently, a structure is also a structure of structures, and a function is a function of functions. From a logical point of view, a system transforms a task into an activity (Figure 3). Note that activity is typically observed as system's behavior.

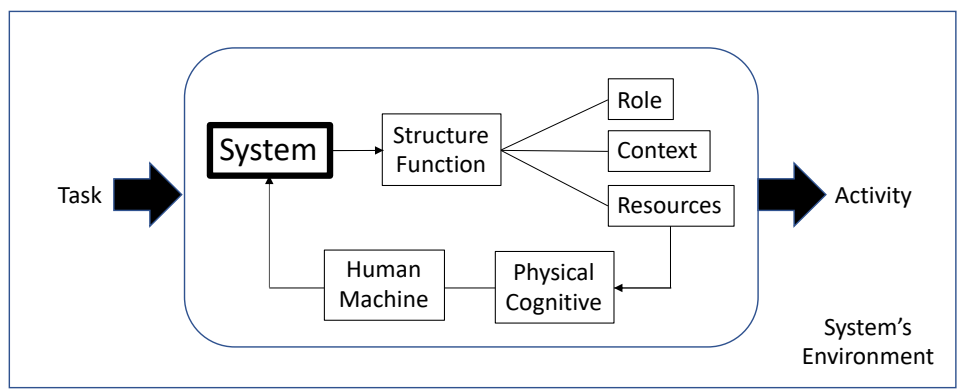

Figure 3. Logical and teleological definition of a system.

It follows that function allocation consists in allocating a function of functions on top of a system of systems. This allocation can be deliberately done at design time, and dynamically at operations time by observing and analyzing system's activity, enabling the discovery of emergent functions and structures. In other words, function allocation can be done incrementally using formative evaluations and agile system development. The way structures and functions are defined and incrementally modified strongly influence system flexibility in terms of both system upgrade and activity adaptation to operational context.

Sociotechnical systems should be *flexible*. First, they should be intrinsically flexible in terms of structures (i.e., the architecture of a system should be designed to possibly enable easy modification of current function allocation) and functions (i.e., a function should be able to perform correctly in a wider context than initially specified). Second, systems should be extrinsically flexible in terms of possibly expended service within the systems of systems where it belongs (i.e., the service that was initially defined can be extended outside the system's context of validity). Intrinsic and extrinsic flexibility of a system involve two crucial properties: adaptability and expendability. They both refer to services. For example, if a machine or human service within a system can be easily replaced by another equivalent human or machine service, then the system will be said to be flexible. This kind of flexibility should exist during the whole life cycle of a (sociotechnical) system at various levels of granularity.

Another property of a sociotechnical system is *separability* (Figure 4). The more complex a system is (the term "complex" is taken in the Latin sense of "complexus," i.e., what is woven together), both structurally and functionally, the more difficult it is, not only to apprehend its behavior, but also to understand its internal functioning (i.e., understanding the interactions between the various components of a system of systems). Can parts of the system be separated to study them in isolation and thus simplify the analysis? This is a difficult question, but one that biologists and physiologists have been considering and studying for a long time. Surgeons, for example, know how to momentarily separate an organ from the human body without irreversibly damaging the whole body, considered as a system of systems. They also know that certain organs, such as the brain, cannot be separated because the human being could die from this separation. These vital organs must be studied and treated while being connected to the rest of the body.

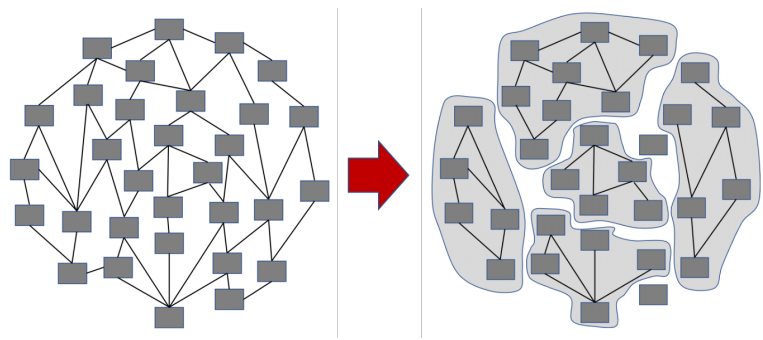

Figure 4. Example of seven separable systems of a system of systems.

The traditional silo engineering design approach associated with late components integration, most often leads to major problems that are difficult, and even impossible, to solve at the end of the chain because rebuilding the entire system is difficult and at times impossible. This approach does not usually consider the separability property and often considers that all system components are separable, like Lego blocks. The more we master the separability property of a system of systems, the more we will be able to address its fluidity and flexibility at operations and maintenance times, for example. It is important to realize that separability was not a major problem if systems were purely mechanical. However, now that systems not only include a large amount of software, but are also highly interconnected, it becomes imperative to understand their intrinsic separability properties. Today, experts solve most problems of these complex sociotechnical systems, often using prosthetic devices, but we need to better formalize them, systemically speaking, to enable finding better and more integrated solutions.

It is interesting to notice that this kind of socioergonomics is based on biological analogs, which refers to Auguste Comte's sociological approach (Comte, 1851–1854; Weinstein, 2019). The range of social scientific methods has also expanded, as social research is nowadays based on a variety of qualitative and quantitative methods. During the second half of the 20th century, societal analyses have been interpretative, hermeneutic, and philosophic (Babich, 2017). Since the beginning of the 21st century, sociological methods are increasingly rigorous analytically, mathematically, and computationally, supported for example by agent-based modeling (Chattoe, 2013) and social network analysis (Borgatti, Everett & Johnson, 2013).

A system of systems is a living entity, which evolves with respect to its activity and accumulated experience. Observed *emergent* behaviors result from an integration of phenomena that involve systems ranging from individual systems to more macroscopic systems (i.e., societies of systems). Resulting emergent behaviors or facts can be interpreted as social facts in sociology (Durkheim, 1895).

## Concluding Remarks

Socioergonomics is strongly based on sociology and related social science disciplines. It is an emerging field that investigates agents, whether humans or machines, in relation to their social behavior and interactions in work environments. Socioergonomics research aims to expand our understanding of the social mechanisms underlying agents' personality, trust, collaboration, individual and collective performance in a real-world sociotechnical environment. This approach can be described as the systemic study of multi-agent mechanisms and sociological structures and functions of humans and machines in work environments. It addresses and attempts to improve knowledge about social processes at work in our contemporary societies, at various levels of granularity (from the system-of-systems viewpoint). It contributes to the development of methods and tools for empirical investigation and critical analysis of such processes. Socioergonomics focuses on group dynamics, social performance, and changes, as well as organizational safety, performance, and well-being.

Socioergonomics expands traditional corrective ergonomics, based on 20th century practice of engineering (i.e., development of engineered systems first and development of user interfaces after). Socioergonomics informs engineering stakeholders during the whole life cycle of a sociotechnical system (STS). It considers the evolution of systems, whether continuously or disruptively, as a holistic endeavor where humans and machines adapt their structures and functions in harmony toward sustainable HSI validity (i.e., economically, environmentally, and societally). To this end, it integrates advancements of social sciences and HCD, to provide meaningful solutions to STSs evolution.

Socioergonomics guides engineering design, refinement, and management of contemporary complex life-critical sociotechnical systems in various industrial and everyday-life domains (e.g., transportation, nuclear, health, defense). It focuses more on longer-term understanding of STSs sustainability (i.e., addressing societal, economic, and environmental issues) and performance (including safety, efficiency and comfort) than on short-term metrics that enable optimization of quick gains. It enables us to study collaborative work environments, shared situation awareness, group decision making, organization design and management, life-critical systems, cooperative and competitive agencies, mutual trust (intersubjectivity), human machine teaming, etc.

## Acknowledgements

The author thanks the members of INCOSE HSI working Group for insightful inputs and debates on HSI during the last few years, as well as the members of FlexTech Chair both at CentraleSupélec and ESTIA, and Caroline Moricot (Panthéon Sorbonne University).

# Biography

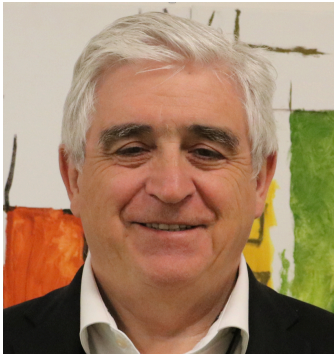

**Guy André Boy,** Ph.D., is a Scientist and Engineer, Fellow of INCOSE and the Air and Space Academy. He is University Professor at CentraleSupélec (Paris Saclay University) and ESTIA Institute of Technology. He was Chief Scientist for Human-Centered Design at NASA Kennedy Space Center, University Professor and Dean at Florida Institute of Technology, where he created the Human-Centered Design Institute. He was Senior Research Scientist at Florida Institute for Human and Machine Cognition, and former President and CEO of the European Institute of Cognitive Sciences and Engineering (EURISCO), France.